\documentclass[aps,prl,superscriptaddress,twocolumn,showpacs,titlepage=false]{revtex4}
\usepackage{graphicx}
\usepackage{amsmath}
\usepackage{amssymb}
\usepackage{subfigure}
\usepackage{natbib}
\usepackage{color}
\usepackage[breaklinks=true,colorlinks=true,linkcolor=blue,urlcolor=blue,citecolor=blue]{hyperref}
\def\bra#1{{\left\langle #1 \right|}}
\def\ket#1{{\left| #1 \right\rangle}}

\newcommand{\bs}[1]{\boldsymbol{#1}}

\begin{document}

\title{Local versus Global Strategies in Multi-parameter Estimation}
\author{P. A. Knott}
\thanks{The first two authors contributed equally to this work}
\affiliation{Department of Physics and Astronomy, University of Sussex, Brighton BN1 9QH, UK}
\author{T. J. Proctor}
\thanks{The first two authors contributed equally to this work}
\affiliation{School of Physics and Astronomy, University of Leeds, Leeds, LS2 9JT, UK}
\affiliation{Berkeley Quantum Information and Computation Center, Department of Chemistry, University of California, Berkeley, CA 94720, USA}
\author{A. J. Hayes} 
\affiliation{Department of Physics and Astronomy, University of Sussex, Brighton BN1 9QH, UK}
\author{J. F. Ralph}
\affiliation{Department of Electrical Engineering and Electronics, The University of Liverpool, Brownlow Hill, Liverpool, L69 3GJ, UK}	
\author{P. Kok}
\affiliation{Department of Physics and Astronomy, University of Sheffield, Sheffield S3 7RH, UK}
\author{J. A. Dunningham}
\affiliation{Department of Physics and Astronomy, University of Sussex, Brighton BN1 9QH, UK}
\date{\today}

\begin{abstract}
We consider the problem of estimating multiple phases using a multi-mode interferometer. In this setting we show that while global strategies with multi-mode entanglement can lead to high precision gains, the same precision enhancements can be obtained with mode-separable states and local measurements. The crucial resource for quantum enhancement is shown to be a large number variance in the probe state, which can be obtained without any entanglement between the modes. This has important practical implications because local strategies using separable states have many advantages over global schemes using multi-mode-entangled states. Such advantages include a robustness to local estimation failure, more flexibility in the distribution of resources, and comparatively easier state preparation. We obtain our results by analyzing two different schemes: the first uses a set of interferometers, which can be used as a model for a network of quantum sensors, and the second looks at measuring a number of phases relative to a reference, which is concerned primarily with quantum imaging.
\end{abstract}
\maketitle

Quantum metrology has the potential to revolutionize a diverse range of fields from biological imaging \cite{taylor2013biological} to navigation \cite{bongs2006high,dowling1998correlated}, and already plays a crucial role in enhancing the precision of gravitational wave detectors \cite{aasi2013enhanced}. In many practical applications it is necessary to estimate multiple parameters \cite{komar2014quantum,freise2009triple,giovannetti2009sub,tsang2009quantum,shin2011quantum}, and hence it is important to understand the potential enhancements that quantum metrology can provide in this setting \cite{humphreys2013quantum,ciampini2015quantum}. It has already been shown that in a multi-mode (multi-path) interferometer, measuring all phases simultaneously with a mode-entangled state can enhance the precision \cite{humphreys2013quantum,baumgratz2015quantum}. However, in stark contrast to this, in other applications of quantum metrology multi-mode entanglement can be detrimental, such as when measuring coupled phases \cite{kok2015quantum} or when loss is considered \cite{knott2014effect}. Furthermore, from a practical point of view large multi-mode-entangled states are notoriously difficult to produce and are fragile to experimental imperfections and photon losses. This warrants further investigation into the role of entanglement and global estimation strategies in multi-mode metrology.

In this paper the problem of multi-parameter estimation in the context of optical interferometry is considered. We show that while multi-mode entanglement and global estimation strategies can lead to high precision gains over standard quantum metrology protocols, the same precision enhancements can also be obtained with mode-separable states and local measurements alone. Local strategies offer a number of advantages over their global counterparts including robustness to local estimation failure, more flexibility in the distribution of resources, and more realistic methods of state preparation \cite{knott2015practical,ourjoumtsev2007generation,huang2015optical,etesse2015experimental,knott2015evolutionary}, measurement and control. Given all these advantages it is interesting to note that multi-mode entanglement is not essential in quantum enhanced optical metrology, and separable pure states with a large number variance can be shown to equal or even surpass their multi-mode-entangled counterparts.

\begin{figure}
\includegraphics[width=7.0cm]{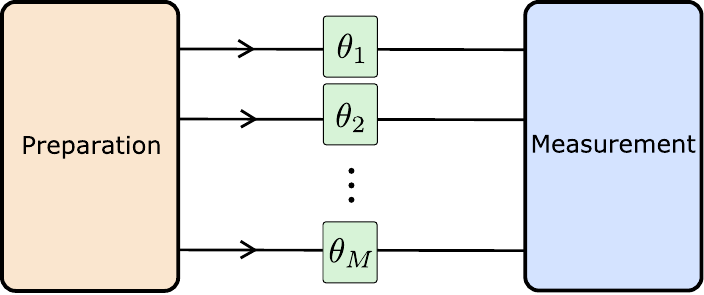}
\caption{The general problem under consideration consists of $M$ optical modes with independent linear phase shifts $\theta_i$, $i=1,...,M$. In optical interferometry the parameters to be estimated, $\phi_i$, are given by some function of the $M$-dimensional vector $\boldsymbol{\theta}$, as described in the main text. For example the $\phi_i$ could be phase differences between arms.}
\label{fig:MPE_general_noref}
\end{figure}

Our results are obtained by analysing two different multi-parameter estimation schemes which cover a variety of practical applications. Firstly, we consider a collection of (possibly entangled) interferometers which can be used as a model for a network of quantum sensors or precision clocks \cite{komar2014quantum}. This scheme is also relevant to applications such as gravitational wave astronomy in which multiple parameters of a gravitational wave will be measured simultaneously \cite{freise2009triple}. Secondly, we analyze a model for quantum-enhanced imaging \cite{giovannetti2009sub,tsang2009quantum,shin2011quantum}, introduced by Humphreys \textit{et. al.} \cite{humphreys2013quantum}, whereby many phases are measured relative to a single global reference. In both these schemes we provide mode-separable states that can surpass their multi-mode-entangled analogues, and we present phase-precision bounds which explicitly show that multi-mode entanglement is not a crucial resource for enhanced metrology.

\textit{Multi-parameter estimation -} Consider the problem of estimating the general vector $\bs{\phi}$ consisting of $d$ parameters $\phi_{i}$, $i=1,...,d$. The precision bound on estimating each parameter $\phi_{i}$ is given by the Cram\'er-Rao bound (CRB) as $\delta\phi_i^2 \geq \mu^{-1} (\mathcal{F}^{-1})_{ii}$ where $\mu$ is the number of repetitions of the experiment and $\mathcal{F}$ is the quantum Fisher information matrix (QFIM) \cite{jarzyna2012quantum,paris2009quantum}. For a pure state $\ket{\psi_{\boldsymbol{\phi}}}$ which depends on $\bs{\phi}$ the QFIM is defined by
\begin{equation}
\mathcal{F}_{l m} = \frac{1}{2} \bra{\psi_{\boldsymbol{\phi}}}( L_{l}L_{m} +L_{m}L_{l})\ket{\psi_{\boldsymbol{\phi}}},
\end{equation}
where $L_{l}$ is the symmetric logarithmic derivative given by $L_{l} = 2(\ket{\partial_{l} \psi_{\boldsymbol{\phi}}} \bra{\psi_{\boldsymbol{\phi}}} +\ket{\psi_{\boldsymbol{\phi}}} \bra{\partial_{l} \psi_{\boldsymbol{\phi}}})$ with $\ket{\partial_{l} \psi_{\boldsymbol{\phi}}}  \equiv \frac{\partial}{\partial \phi_l} \ket{\psi_{\boldsymbol{\phi}}}$ \cite{humphreys2013quantum,helstrom1976quantum}. Consider the case when $\ket{ \psi_{ \bs{\phi} } }= U(\bs{\phi})\ket{\psi}$ for some $\bs{\phi}$-independent initial probe state $\ket{\psi}$ and $U(\boldsymbol{\phi}) = \exp ( i \sum_{i =1}^{d} \phi_{i} \hat{O}_{i} )$ where the $\hat{O}_{i}$ are Hermitian and mutually commuting operators, i.e. $[ \hat{O}_i,\hat{O}_j]=0$ $\forall i,j$. Then it can be shown that $\mathcal{F}_{lm} =  4 \text{Cov}(\hat{O}_l ,\hat{O}_m)$,
where $\text{Cov}(\hat{O}_l,\hat{O}_m) = \langle \hat{O}_l\hat{O}_m \rangle -  \langle \hat{O}_l\rangle \langle \hat{O}_m \rangle$ is the covariance between the two operators $\hat{O}_l$ and $\hat{O}_m$, and the expectation values are taken with respect to the input state $\ket{\psi}$ (this simple QFIM formula will  be applicable throughout). The variance, given by $l=m$, will be denoted $\text{Var}(\hat{O}_l)=\text{Cov}(\hat{O}_l,\hat{O}_l)$. The general scheme for optical multi-parameter estimation considered herein is shown in Fig.~\ref{fig:MPE_general_noref}. There are $M$ optical modes with independent linear phase shifts. The unknown phase shifts are imprinted with the unitary operator $U(\boldsymbol{\theta}) = \exp ( i \sum_{j=1 }^{M} \theta_{j} \hat{n}_{j})$ and the problem is to estimate some number $d \leq M$ of independent parameters $\phi_i$ which are functions of the $\theta_j$, as will become clear when we introduce specific examples below.

\begin{figure}
\centering
\includegraphics[scale=1.0]{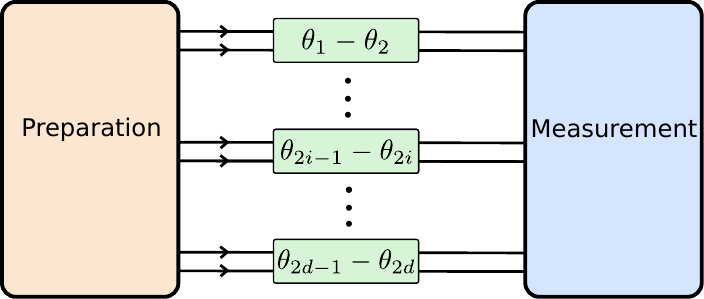}
\caption{A network of quantum sensors may be modelled as $d$ parallel interferometers. The parameters to be measured are the phase differences in each interferometer.}
\label{fig:d_interferometers}
\end{figure}

\textit{Parallel interferometers -} The first scheme we consider is a set of parallel interferometers in which the aim is to measure the phase difference between the two arms in each interferometer, as shown in Fig.~\ref{fig:d_interferometers}. One interesting future application is in gravitational wave astronomy, which will aim to simultaneously measure a number of parameters associated with gravitational waves, such as polarisation and direction of origin, and to do so will require multiple interferometers \cite{freise2009triple}. This scheme can also model quantum sensing networks such as networks of precision clocks, as proposed in \cite{komar2014quantum}.

This parallel interferometers model is a special case of the scheme in Fig.~\ref{fig:MPE_general_noref} for an even number of modes $M=2d$, where specifically we take the $i$\textsuperscript{th} interferometer to consist of modes $2i -1$ and $2i$ ($i=1, \dots, d$). The aim is to estimate the $d$ parameters $\phi_i\equiv \phi_{i^{-}}$ where $\phi_{i^{\pm}} = \theta_{2i-1} \pm \theta_{2i} $. The phase-shift operator $U(\boldsymbol{\theta}) $ can be re-parameterised in terms of $\boldsymbol{\phi}=(\phi_{1^-},\dots,\phi_{d^-},\phi_{1^+},\dots,\phi_{d^+})$, giving
$
U(\boldsymbol{\phi})  = \exp ( i \sum_{i=1}^{d} (\phi_{i^-} \hat{O}_{i^-}+ \phi_{i^+} \hat{O}_{i^+} )),
$
where the generating operators are  $\hat{O}_{i^{\pm}} =    (\hat{n}_{2i-1} \pm \hat{n}_{2i})/2$.
Hence, although the estimation is only of $d$ parameters, the relevant QFIM is for the $2d$-dimensional $\boldsymbol{\phi}$ and has the form $\mathcal{F}_{i^{\pm} j^{\pm} }  = 4 \text{Cov} ( \hat{O}_{i^{\pm}} ,\hat{O}_{j^{\pm}} )$,
where the two $\pm$ signs may be chosen independently. 

This estimation problem has a symmetry between the interferometers and, furthermore, there is a symmetry between the arms in each interferometer as neither plays a special role. We therefore consider states that are symmetric with respect to swapping interferometer labelling, and symmetric with respect to swapping the modes in each interferometer. Using the shorthand $C_{i,j} \equiv \text{Cov} ( \hat{n}_{i} ,\hat{n}_{j} )$ and $V_{i} \equiv \text{Var} ( \hat{n}_{i} )$, these symmetry assumptions imply that the variances of all the modes are equal, i.e., $V_{i} = V_j$ for all  $i$ and $j$ and this value may be denoted $V$. Furthermore, they imply that the covariances between any two modes from the same interferometer are equal, i.e. $C_{2i-1,2i} =C_{2j-1,2j}$ for all $i$ and $j$ and this value may be denoted  $C_{\text{Intra}}$. Given these natural symmetries it can be shown (see Appendix A) that the precision bound for estimating each parameter $\phi_i$ is given by
\begin{align}
\label{eq:prec_parallel_int}
\delta \phi_i^2 \geq \frac{1}{2(V - C_{\text{Intra} } )}.
\end{align}
In the literature a single phase-precision parameter $\delta \Phi = \sum_{i=1}^{d} \delta \phi_i$ is sometimes considered, e.g., see \cite{humphreys2013quantum,liu2014quantum}, which here may be trivially calculated to be $\delta \Phi = d \delta \phi_i $, but throughout this paper we will consider the precision bounds of individual phases $\delta \phi_i $. From Eq.~(\ref{eq:prec_parallel_int}) is clear that the only parameters which directly affect the phase precision are the state's photon number variance and the correlations between the two modes in an individual interferometer. Hence, entanglement between interferometers provides no direct improvement in the phase precision. It is therefore not necessary to entangle quantum optical sensors in networks, nor entangle multiple gravitational wave interferometers, which in both cases would be challenging.

It is instructive to rewrite Eq.~(\ref{eq:prec_parallel_int}) in terms of the Mandel $\mathcal{Q}$ parameter and the two-mode correlation parameter $\mathcal{J}_{ij}$, which are defined by $\mathcal{Q}_i =  ( V_i - \bar{n}_i )/ {\bar{n}_i}$ and $\mathcal{J}_{ij} = {C_{i,j}}/{\sqrt{V_i V_j}}$ respectively. We denote the Mandel-$\mathcal{Q}$ parameter for any mode by $\mathcal{Q}$ (all modes have the same $\mathcal{Q}$), and the two-mode correlation between the two modes in any of the interferometers by $\mathcal{J}$, where $\mathcal{J} = C_{\text{Intra}}/V$. Then for all $i$ the phase precision is given by:
\begin{align}
\delta \phi^2_i \geq  \frac{1}{2\bar{n} (1+\mathcal{Q} )(1 - \mathcal{J} )} ,
\label{parallel_pre}
\end{align}
where $\bar{n}$ is the average number of particles in any \emph{single mode} (i.e. $\bar{n} = \bar{n}_i = \langle \hat{n}_i \rangle$ for any $i$). For single-parameter estimation this was shown in Ref. \cite{sahota2015quantum}. This may also be rewritten in terms of the average \emph{total} photon number $\bar{N}$ using $\bar{N}=2d\bar{n}$. The 2-mode correlation term is bounded by $-1 \leq \mathcal{J} \leq 1$ and hence only provides at most a factor of $1/\sqrt{2}$ improvement in the phase precision.

We now compare local and global phase estimation strategies with examples of both multi-mode-entangled and mode-separable states. If we consider each interferometer individually, the standard quantum-enhanced precision is the well-known Heisenberg scaling of
$\delta \phi^2_i \geq 1/(2\bar{n})^2 =  d^2/\bar{N}^2$, where this precision for each individual phase has been now written in terms of the total photon number $\bar{N}$ used to measure all of the phases. Consider a generalised entangled coherent state (GECS) given by
\begin{equation}
\ket{\Psi_{\textsc{gecs}}} = \mathcal{N}_g \sum_{a \epsilon \mathcal{M}} \hat{D}_a(\alpha_g)\ket{\textbf{0}},
\end{equation}
where $\hat{D}_a(\alpha) = \exp{ (\alpha \hat{a}^{\dagger} - \alpha^* \hat{a}) }$ is the displacement operator acting on mode $a$, $\mathcal{M}$ is the set of $M=2d$ modes, $\ket{\textbf{0}}$ is the multi-mode vacuum state and $\mathcal{N}_g$ is a normalisation factor required due to the non-zero overlap of a coherent state with the vacuum. We find
\begin{equation}
\delta\phi_{\textsc{gecs}}^2 \geq \frac{d}{\bar{N}_{g} \left(|\alpha_g|^2 +1\right)}  \approx \frac{d}{\bar{N}_g  \left(\bar{N}_g+1 \right)},
\end{equation}
where  $\bar{N}_g = |\alpha_g|^2 / (1+(2d-1)e^{-|\alpha_g|^2})$ is the total average number in the GECS and the approximation uses $\bar{N}_g \approx |\alpha_g|^2$ which holds for $|\alpha_g| \gg 1$. This is a scaling of $\mathcal{O}(d/\bar{N}_g^2)$ which is an $\mathcal{O}(d)$ improvement over the expected quantum enhancement. This suggests that contrary to the evidence of Eq.~(\ref{parallel_pre}), a global strategy does provide an improvement over the local estimation strategy. However, a more optimal implemenation using a local strategy can do just as well or even better as we will now see.


Consider a multi-mode but \emph{mode-separable} unbalanced cat state (UCS), given by $\ket{\Psi_{\textsc{ucs}}} =  \mathcal{N}_c \left(\ket{\alpha_c} + \nu |0\rangle \right)^{\otimes 2d}$
where $\nu$ is a real parameter and again $\mathcal{N}_c$ is the normalisation. We find that
\begin{equation}
\delta\phi_{\textsc{ucs}}^2 \geq \frac{d}{\bar{N}_c \left(|\alpha_c|^2+1-{\bar{N}_c \over 2d} \right)} 
 \approx \frac{d}{\bar{N}_c \left( {\nu^2 \over 2d} \bar{N}_c +1\right)},
\label{pre_UCS}
\end{equation}
where $\bar{N}_c =2d|\alpha_c|^2/(\nu^2+1+2\nu e^{-\frac{1}{2}|\alpha_c|^2})$ is the total average photon number and the approximation is for $|\alpha_c| \gg 1$. For $\nu=1$ (an ordinary cat state) we find that the precision bound scales as $ \mathcal{O}(d^2/\bar{N}_c^2)$, as perhaps expected of the local strategy. However, if we instead take $\nu^2$ to scale with $d$ then it has the form $\mathcal{O}(d/\bar{N}_c^2)$. More explicitly, setting photon numbers equal $\bar{N}_c = \bar{N}_g$, then (for $|\alpha_c|,|\alpha_g| \gg 1$) we have $\delta\phi_{\textsc{ucs}}^2 < \delta\phi_{\textsc{gecs}}^2$ when $\nu^2 > 2d$ (this analysis also holds without taking the large photon number limit). This shows that for large enough values of $\nu$ the UCS can attain a better precision than the GECS. Before further discussion on the implications of these results, we will now show that similar conclusions can be drawn for an alternative `quantum imaging' problem.

\textit{Multi-mode quantum-enhanced imaging -} Consider measuring $d$ phase shifts relative to a single global reference mode, as described by Humphreys \emph{et. al.} \cite{humphreys2013quantum}, which is relevant for a range of applications, including quantum enhanced imaging \cite{giovannetti2009sub,tsang2009quantum,shin2011quantum}. This is again a special case of Fig.~\ref{fig:MPE_general_noref} for $M=d+1$ modes, whereby the aim is to estimate the $d$ dimensional vector parameter $\boldsymbol{\phi}$ where $\phi_i = \theta_i - \theta_{d+1}$. For simplicity (and following Humphreys \emph{et. al.} \cite{humphreys2013quantum}) we set $\theta_{d+1}=0$ in which case the generator of $\phi_i$ is simply $\hat{n}_i$, and therefore $\mathcal{F}_{ij} = 4 \text{Cov} ( \hat{n}_i, \hat{n}_j)$. As in the case of the parallel interferometers, there is a clear symmetry to this problem, and in this case it is natural to assume symmetry between the $d$ probe modes (but not necessarily between the reference mode and the others). This implies that $V_i = V_j$ for all $i$ and $j$, which is denoted $V$, and that $C_{i,j} = C_{m,n}$ for all $i \neq j $ and $m \neq n$, which we denote by $C$. Using this assumption, it is shown in Appendix B that the precision bound for estimating each parameter $\phi_i$ is given by
\begin{equation} \delta \phi_i^2 \geq  \frac{V+(d-2)C  }{4(V- C )(V+(d-1)C)}. \end{equation}
Again, the QFIM can be expressed in terms of the Mandel-$\mathcal{Q}$ parameter of any mode and the two-mode correlation $\mathcal{J}=C/V$, which gives a phase-precision of
\begin{align}
\delta \phi_i ^2 &\geq   \frac{ f(d,\mathcal{J}) }{4\bar{n} (1+\mathcal{Q} )(1-\mathcal{J}) }  .
\end{align}
where $\bar{n}$ is the average photon number in a single mode and the function $ f(d,\mathcal{J})$ is given by
\begin{equation} f(d,\mathcal{J}) = \frac{1+(d-2)\mathcal{J}}{1+(d-1)\mathcal{J}}. \end{equation}
When there are many interferometers ($d \gg 1$) then $f(d,\mathcal{J}) \approx 1$, and hence the phase precision has a very similar form to that for the parallel interferometers case given in Eq.~(\ref{parallel_pre}). As always, $|\mathcal{J}| \leq 1$ and hence as before multimode correlations can only provide a small constant factor improvement.

In order to explore this further and to understand the relationship to previous work \cite{liu2014quantum,humphreys2013quantum}, examples are now considered. Humphreys \emph{et. al.} \cite{humphreys2013quantum} introduced the generalised NOON state (GNS), given by
\begin{multline*}
\label{eq:unbal_GNS}
\ket{\Psi_{\textsc{gns}}} = \frac{1}{\sqrt{d+\gamma^2}}\big( \ket{N,0, \dots,0,0} + \ket{0,N,\dots,0,0} + \dots \\ + \ket{0,0,\dots,N,0}  +  \gamma \ket{0,0,\dots,0,N} \big).
\end{multline*}
where the real parameter $\gamma$ is a weighting on the reference mode to be optimised. For each phase, the precision bound is $\delta \phi_{\textsc{gns}}^2 \geq (d+\gamma^2)(1+\gamma^2)/4 \gamma^2N^2$,
which is optimised for $\gamma = d^{1/4}$ but for which the simpler choice of $\gamma=1$ provides the same scaling enhancement. The optimal case gives $\delta \phi_{\textsc{gns}}^2 \geq (1+\sqrt{d})^2/4N^2$. This is an $\mathcal{O}(d)$ enhancement over the expected quantum enhancement \cite{humphreys2013quantum}  (separate NOON states give a precision $\delta \phi_{\textsc{noon}}^2 = d^2/4N^2$) which again suggests that a global strategy does provide an improvement over the local estimation strategy. However, a collection of single-mode unbalanced `NO' (UNO) states $\ket{\psi_{\textsc{uno}}} =\mathcal{N}_{_\textsc{uno}} \left(   \ket{N} + \nu \ket{0}\right)$ may again be used to equal or surpass this phase estimation precision by tuning $\nu$. Choosing $\nu = 1$ returns the same scaling as using separate NOON states. However, if we take $\nu =\sqrt{d+\gamma^2-1}$, or simply $\nu \propto \sqrt{d}$, then we obtain exactly the same precision scaling enhancement as the global estimation strategy with the GNS. Furthermore, the multi-mode correlations in the GNS die off with increasing $d$, as $\mathcal{J}=-1/(d+\gamma^2-1)$.

It is now clear that, for quantum-enhanced optical multi-parameter estimation, the essential property required of a pure probe state is large correlations \textit{within} each mode, and this can be obtained without multi-mode entanglement. The cause of the apparent scaling improvement for the global strategy is that the GNS exhibits the scaling $\mathcal{Q} = \mathcal{O}(d\bar{n}) =  \mathcal{O}(\bar{N})$ rather than  $\mathcal{Q} = \mathcal{O}(\bar{n})$, i.e.  the uncertainty in the photon number of each mode grows with the number of modes $d$, for fixed $\bar{n}$. However, the $\mathcal{Q}$ function is simply a local property of each mode, and the desired scaling can also be obtained by a judicious choice of a single-mode state.

Generally, for any path-symmetric pure state of $M$ modes $\ket{\Psi}$, consider a pure single-mode state $\ket{\psi(\Psi)} = \sum_{n=0}^{\infty} | \langle n \ket{\Psi }| \ket{n}$, with $ \langle n \ket{\Psi }$ taken with respect to any mode. Then, by construction, $\ket{\Psi}$ and the $M$-mode-separable state $\ket{\psi}^{\otimes M}$ contain the same average number of photons and for any mode $\mathcal{Q}(\ket{\Psi}) = \mathcal{Q}(\ket{\psi}^{\otimes M})$. Hence the phase-precision as a function of $\bar{n}$ (in either scenario considered herein) for a general multi-mode state exhibits at most a small constant factor (at best $\sqrt{2}$) improvement over the separable analogue. This argument applies to any global estimation strategy, and hence to the extension of Ref. \cite{humphreys2013quantum} by Liu \emph{et. al.} \cite{liu2014quantum} to quantum imaging with a generalised entangled coherent state (GECS).

\emph{Discussion -} We have shown that in optical multi-parameter estimation there is no fundamental improvement in using a global strategy to estimate all of the parameters simultaneously. Local strategies are just as effective, and this has important practical implications because local estimation strategies, which use separable states and local measurements, have a number of advantages. For example, local strategies have greater flexibility in the distribution of resources and are more robust to local estimation failure and errors in state preparation. Furthermore, single mode states with a large number variance can be made in experiments \cite{knott2015practical,ourjoumtsev2007generation,huang2015optical,etesse2015experimental}, and realistic schemes have been proposed to produce separable states which improve over the shot noise limit by more than a factor of 4 \cite{knott2015evolutionary}. By comparison, multi-mode-entangled states with large photon numbers are notoriously difficult to make -- the largest two-mode optical NOON state that has been made experimentally contains only 5 photons \cite{afek2010high}.




We note that the QFI alone is not always a reliable method for deriving precision scaling bounds that are truly attainable in practice, and a proper consideration of the prior information and the required number of experimental repetitions is needed. Indeed, states with arbitrarily large QFI for a fixed number of photons have been reported in the literature \cite{rivas2012sub}, and this effect is relevant here. A further discussion of this is given in Appendix C. However, the precision \emph{scaling} with photon number is often not of direct relevance in an experiment, and a more relevant measure is the absolute precision that can be obtained given an allowed total photon number through the interferometer \cite{wolfgramm2013entanglement,taylor2015quantum,aasi2013enhanced}. Our results here still imply that there is no good reason to attempt a global estimation strategy to achieve an improved absolute precision, and local strategies are preferred. As already noted, there are a range of practical states which improve on the absolute precision of NOON states \cite{knott2015practical,knott2015evolutionary}, and these are candidates for the multi-parameter paradigm using a local estimation strategy considered herein.

To conclude, we have considered the problem of multi-parameter estimation in optical interferometry, and shown that local estimation strategies using separable states can surpass the precision enhancements attained by global estimation using multi-mode entanglement. These results hold for quantum sensing, in which a number of phases are measured relative to a reference, and also for a set of parallel interferometers, which can serve as a model for a network of sensors. Local strategies offer many practical advantages over their global counterparts, including flexibility, practicality and control, and therefore should be considered as the preferred method for multi-parameter estimation.

\subsection{Acknowledgements}
We thank Animesh Datta for helpful suggestions. This work was partly funded by the UK EPSRC through the Quantum Technology Hub: Networked Quantum Information Technology (grant reference EP/M013243/1).

\subsection{References}


\bibliographystyle{apsrev}
\bibliography{MyLib_Thesis_multiparameter}


\label{sec:apx}

\subsection{Appendix A: Phase precision derivation for the parallel interferometers}

We begin with the quantum Fisher information matrix (QFIM) equation from the main text:
\[
\mathcal{F}_{i^{\pm} j^{\pm} }  = 4 \text{Cov} ( \hat{O}_{i^{\pm}} ,\hat{O}_{j^{\pm}} ), 
\]
where $\hat{O}_{i^{\pm}} =    (\hat{n}_{2i-1} \pm \hat{n}_{2i})/2$ and consider simplification under the assumption that the input state is symmetric both between and inside interferometers. 
As interferometer $i$ consists of modes $2i-1$ and $2i$, these assumptions imply that
\begin{align*} 
&\forall i\neq j  \hspace{0.5cm} C_{\text{Intra}} \equiv  C_{2i-1,2i} =C_{2j-1,2j}\\
&\forall i,j  \hspace{0.8cm}  V \equiv V_{i} = V_j 
  \end{align*}
  using the short-hand introduced in the main text that $C_{i,j} \equiv \text{Cov}(\hat{n}_i,\hat{n}_j)$ and $V_{i} \equiv \text{Var}(\hat{n}_i)$. These equalities state that the covariances between any two modes from the same interferometer are equal and the variances of all the modes are equal. An implication of the symmetry assumptions is that
\[C_{2i-1,j} = C_{2m-1,n}\]
 whenever $j\neq 2i-1,2i$ and $n \neq 2m-1,2m$ (as a covariance is symmetric this covers all remaining cases), and this value may be denoted $C_{\text{Inter}}$ as it represents any correlations between interferometers. Note that total path symmetry can be enforced by letting $C_{\text{Intra}}=C_{\text{Inter}}$, but there is no need to make this assumption (and it is not automatically sensible given the symmetry of the problem).

The QFI is now simplified under these assumptions. The elements of the QFI matrix can be expanded to
\begin{multline*} \mathcal{F}_{i^{\pm_p}j^{\pm_q} } = C_{2i-1,2j-1}  (\pm_p )(\pm_q) C_{2i,2j} \\ 
\pm_p C_{2i,2j-1} \pm_q C_{2i-1,2j},\end{multline*}
where the subscripts $p$ and $q$ on the $\pm$ symbols are used to explicitly denote that these may be taken to be $+$ or $-$ independently, and the subscripts show how they match together. Using the assumptions above it is easily confirmed that
\[
 \mathcal{F}_{i^{\pm}j^{\mp} }  =0 ,\hspace{0.5cm} \mathcal{F}_{i^{-}j^{-} }  =0 \hspace{0.5cm} i\neq j .
 \]
The final terms for $i \neq j$ are all equal and given by
\[
 \mathcal{F}_{i^{+}j^{+} }  = 4C_{\text{Inter}}.
 \]
Consider then $i=j$. We have that
\[ 
\mathcal{F}_{i^{\pm_p}i^{\pm_q} }  = V  (\pm_p )(\pm_q)  V \pm_p  C_{\text{Intra}} \pm_q C_{\text{Intra}}.
\] 
Hence, all of the diagonal terms are one of two values given by
\[  
\mathcal{F}_{i^{\pm}i^{\pm} } = 2  ( V \pm C_{\text{Intra}}) , \hspace{0.5cm} \forall i.
\]
Finally, it is clear that $\mathcal{F}_{i^{\pm}i^{\mp} }=0$. Hence, combining all of these terms into the QFIM gives
\[
\mathcal{F}  = \begin{pmatrix}  2  ( V - C_{\text{Intra}}) \mathbb{I} & 0  \\ 0 & M \end{pmatrix}, 
\]
where $\mathbb{I}$ is the $d \times d$ identity matrix and $M =  \lambda (\mathbb{I} + \omega \mathcal{I}) $ where $\lambda =  2(V + C_{\text{Intra}}-2C_{\text{Inter}})$, $\omega = 2 C_{\text{Inter}} / (V + C_{\text{Intra}}-2C_{\text{Inter}})$ and $\mathcal{I}$ is the $d \times d$ matrix of all ones. The inverse of any matrix with the form of $M$ is given by
 \begin{equation}
 M^{-1} = \frac{1}{\lambda} \left( \mathbb{I} - \frac{\omega}{1+\omega d} \mathcal{I} \right),
 \label{Eq:M-inverse}
 \end{equation}
 as may be easily confirmed directly by noting that $\mathcal{I}^2 = d \mathcal{I}$. However, we are not actually interested in these terms (the parameters of interest are $\phi_i \equiv \phi_i^-$, we are not attempting to also estimate the $\phi_i^+$). The inverse of $\mathcal{F}$ may then simply be written as
\begin{align} \mathcal{F}^{-1}  = \begin{pmatrix} \frac{1}{2(V_s - C_{ss'})} \mathbb{I} & 0  \\ 0 & M^{-1} \end{pmatrix}. \end{align}
This gives the phase precision bound for the terms of interest (the $\phi_i$) to be
\[
\delta \phi_i^2 \geq \frac{1}{2(V - C_{\text{Intra} } )},
\]
as stated in the main text. Note that this is independent of $d$ and as required it agrees with the single parameter (i.e., single-interferometer) estimation case ($d=1$), e.g., see Ref. \cite{knott2015practical}.

\subsection{Appendix B: Phase precision derivation for quantum imaging}

We begin with the QFIM from the main text $\mathcal{F}_{ij} = 4\text{Cov} (\hat{n}_{i}, \hat{n}_j) =4 C_{i,j}$.
The assumption of path-symmetry between the $d$ (probe) modes, as stated in the main text, implies that $V_i = V_j$ for all $i$ and $j$, which is denoted $V$, and that $C_{i,j} = C_{m,n}$ for all $i \neq j $ and $m \neq n$, which we denote by $C$. Then it immediately follows that $\mathcal{F}_{ii} = 4 V$ for all $i$ and $\mathcal{F}_{ij} = 4 C$ for all  $i\neq j$.
Hence the QFI matrix may be written in the form
\[ \mathcal{F} = 4(V- C) \left( \mathbb{I}+ \frac{C}{V- C}  \mathcal{I} \right), \]
where again $\mathcal{I}$ and $\mathbb{I}$ are the $d \times d$ matrix of all ones and the identity respectively. The inverse of such a matrix is given in Eq.~(\ref{Eq:M-inverse}), and using this formula we have
\[
 \mathcal{F}^{-1} = \frac{1}{4(V- C)} \left( \mathbb{I} -  \frac{C}{V +(d-1)C } \mathcal{I} \right).
  \]
This then implies that for all $i$, the phase precision bound for $\phi_i$ is
\[
\delta \phi^2_{i}  \geq   \frac{V+(d-2)C}{4(V- C )(V+(d-1)C)}.
 \]
as stated in the main text. Note that for the single parameter estimation case ($d=1$) this reduces to $1/4V$ as expected.

\subsection{Appendix C: The QFI as a figure of merit}

The QFI alone is not always a reliable method for deriving precision scaling bounds that are truly attainable in practice. In general, the precision as obtained by the Cram\'er-Rao bound (CRB), $\delta\phi_i^2 \geq \mu^{-1} (\mathcal{F}^{-1})_{ii}$, is achievable given a certain level of prior knowledge of the phase, and an asymptotically large number of repetitions, $\mu$. Indeed, the unbalanced cat state (UCS), given in the main text by $\ket{\Psi_{\textsc{ucs}}} =  \mathcal{N}_c \left(\ket{\alpha_c} + \nu |0\rangle \right)^{\otimes 2d}$, has already been considered in optical quantum metrology and, as shown in \cite{rivas2012sub}, it has an unbounded precision for fixed $\bar{n}$. This can be seen by considering Eq.~(6) in the main text and allowing $\nu$ to grow without bound. The root of this strange effect is that the QFIM is a measure of how a probe state transforms with an infinitesimal change in the parameter to be estimated and does not take into account any further important details such as the level of prior knowledge required of each phase, or the number of experimental repetitions required to obtain this precision. For single-parameter estimation, it is known that states such as the UCS cannot in practice provide a ``sub-Heisenberg'' scaling \cite{hall2012universality,giovannetti2012sub}. These results have been extended to the multi-parameter case, and it has been shown that a sub-Heisenberg scaling cannot be achieved here either \cite{zhang2014quantum,tsang2014multiparameter}. Despite this, the scaling with photon number is often not of direct relevance in an experiment, and a more relevant measure is the absolute precision that can be obtained given an allowed total photon number through the interferometer \cite{wolfgramm2013entanglement,taylor2015quantum,aasi2013enhanced}. In single parameter estimation, squeezed cat states, which have a large $\mathcal{Q}$, have recently been shown to obtain an improved absolute precision over NOON states \cite{knott2015practical}, and we expect these results to be applicable in the multi-parameter case.




\end{document}